\documentclass[10pt,letterpaper,twocolumn]{article}
\usepackage{ol}
\usepackage[ansinew]{inputenc}

\usepackage{graphicx}           
\usepackage{amsmath,amssymb}

\newcommand{\iea}[0]{{\it et al.}}

\begin{document}

\twocolumn[
\title{Adiabatic frequency conversion of optical information in atomic vapor}

\author{Frank Vewinger\cite{vewinger}, J\"urgen Appel, Eden Figueroa, A. I. Lvovsky}

\affiliation{Institute for Quantum Information Science, University of
  Calgary, Calgary, Alberta T2N 1N4,
  Canada}
\homepage{http://www.iqis.org/}

\date{\today}

\begin{abstract}
We experimentally demonstrate a communication protocol that enables frequency conversion and routing of quantum information in an adiabatic and thus robust way. The protocol is based on electromagnetically-induced transparency (EIT) in systems with multiple excited levels: transfer and/or distribution of optical states between different signal modes is implemented by adiabatically changing the control fields. The proof-of-principle experiment is performed using the hyperfine levels of the rubidium D1 line.

\end{abstract}
]


An essential element of a quantum optical communication network is a tool for transferring and/or distributing quantum information between optical modes (possibly of different frequencies) in a loss- and decoherence-free fashion. This is important not only for routing quantum information, but also for interfacing quantum communication lines of different wavelength (e.g. fiber-optical and open-air) between each other and with memory-based quantum repeaters \cite{DLCZ,lightstorage}.

Some experiments on frequency conversion of quantum states of light have been performed using nonlinear optical effects in crystals \cite{Tanzilli:05,huang:92,giorgi:03} and periodically-poled waveguides \cite{upconversion,kwiat04}. An alternative approach\cite{Zibrov:02,wang:05} involves storage of light by means of EIT \cite{fleischhauer:633} and its subsequent retrieval on another optical transition.


In this paper, following our group's recent proposal \cite{appel:06}, we experimentally demonstrate a protocol for routing and frequency conversion of optical quantum information via EIT in an atomic system with multiple excited levels. Our method is related to that of Zibrov \iea\cite{Zibrov:02} and Wang \iea\cite{wang:05}, but  here the information is transferred \emph{during the propagation}, thus avoiding the losses associated with the storage of light. By means of the EIT control fields we steer the composition of the optical component of the dark-state polariton, and can convert, completely or partially, the incoming signal state into another optical mode.

Our scheme (that we call RATOS, Raman adiabatic transfer of optical states) resembles stimulated Raman adiabatic passage (STIRAP) \cite{STIRAP}, but applies to optical rather than atomic states.
Thanks to its adiabatic character, the efficiency of RATOS does not strongly depend on the parameters of the control fields, but only on their initial and final values, which is favorable for possible practical applications of the method.

\begin{figure}[t]
\centerline  {\includegraphics[width=0.9\columnwidth]{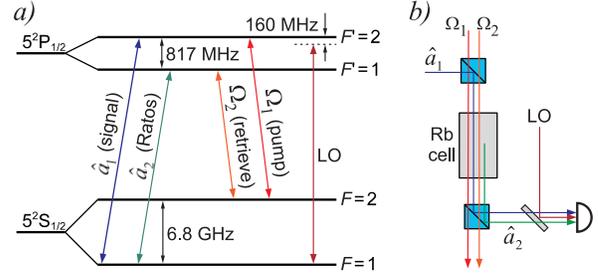}}
  \caption{\small a) Transitions used in the experiment, as described in the text. Also shown is the local oscillator field (LO) used for heterodyne detection.
b) A sketch of the experimental setup. In the actual experiment, the beams are
  overlapping in the cell, the separation in the drawing is for clarity. }\label{fig:levelscheme}
\end{figure}

We consider a double-$\Lambda$ system as shown in Fig.~\ref{fig:levelscheme}(a), with the rubidium D1 transition (where the ground $5S_{1/2}$ and excited $5P_{1/2}$ levels are each split into two hyperfine sublevels) as a model. The energy levels are coupled by two weak signal fields, described by their annihilation
operators $\hat a_1$ and $\hat a_2$, and two strong control fields, described by their Rabi frequencies $\Omega_1$ and $\Omega_2$. Such a system
exhibits EIT for the superposition
\begin{equation}\label{superposition}
    \hat b \propto \left[\frac{\Omega_1}{g_1}\hat a_1
    + \frac{\Omega_2}{g_2}\hat a_2\right],
\end{equation}
of the signal fields, with $g_i$ being the vacuum Rabi frequency for the $i$th signal mode\cite{Liu06,appel:06,Moiseev}.

RATOS proceeds as follows. With only control field 1 (hereafter called pump) initially present, a pulsed optical state in mode $\hat a_1$ (signal) is coupled into the medium. While it is propagating, control field
2 (retrieve) is turned
on slowly, so the EIT signal mode is adiabatically converted
into a superposition (\ref{superposition}), which continues to propagate
losslessly through the cell \cite{appel:06}.

If the pump field is left on, the optical state that has entered the cell in mode  $\hat a_1$ will leave it in mode $\hat b$. This allows the implementation
of beam splitting for optical modes of different frequency, with the final intensities
of the two control pulses determining the outcome of the
process. The beam splitting ratio is
given by
\begin{equation}\label{splittingratio}
    \frac{\langle\hat a_2\rangle}{\langle\hat a_1\rangle} =
    \frac{g_1\Omega_2}{g_2\Omega_1}\,.
\end{equation}
If the pump field is instead adiabatically switched off while mode $\hat b$ is still inside the cell, the initial quantum state of mode $\hat a_1$ will be fully transferred to the Ratos mode $\hat a_2$, thus completing the RATOS protocol.



\begin{figure}[t]
 \centerline{\includegraphics[width=0.9\columnwidth]{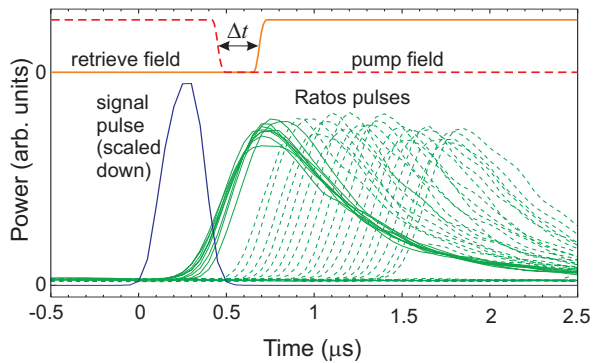}}
  \caption{\small Temporal profiles of the fields. Ratos pulses are shown for the case the retrieved field is turned on before (solid lines) and after (dashed lines) the pump field has been turned off.}\label{fig:pulses}
\end{figure}

\begin{figure}[b]
\centerline{\includegraphics[width=0.9\columnwidth]{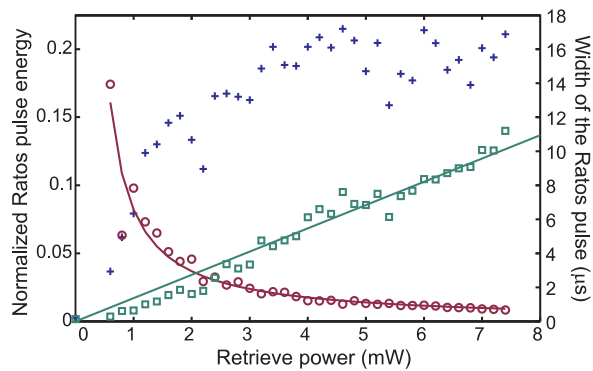}}
  \caption{\small Peak power of the retrieved Ratos pulse ($\square$), its
  temporal width ($\circ$), and energy ($+$) as a function of the power of the retrieve laser.
  The energy is normalized to the energy of the slowed down pulse. The peak intensity plot is in arbitrary units. The solid lines are a linear and inverse linear fits.}\label{fig:powerdependence}
\end{figure}

The experiment was performed in warm
rubidium-87 vapor at 60$^{\circ}$C in a 5-cm long cell filled with 5
Torr of neon as a buffer gas. The cell was mounted within a
magnetically shielded oven. 
The signal field was provided by a
Coherent MBR-110 Ti:Saphire laser with a narrow spectral width
($\sim$40 kHz) and high long-term stability. The pump and retrieve fields were obtained from external-cavity diode
lasers sequentially phase-locked to each other and the Ti:Sapphire laser. Both phase lock circuits were programmed to ensure that the frequency difference among the three fields entering the cell corresponded to the hyperfine splitting frequencies of the ground and excited levels of the rubidium D1 transition: 6835 and 817 MHz, respectively. All three laser beams were controlled by acousto-optical modulators. All polarizations were linear, with the polarization of the control beams orthogonal to that of the
signal and Ratos beams. 

After passing through the cell, the
two signal fields were separated from the control beams using a
polarizing beamsplitter, and subjected to heterodyne detection on a fast photodiode [Fig.~\ref{fig:levelscheme}(b)]. The role of the local oscillator was played by the unmodulated field of the Ti:Sapphire laser. The beat note at 160~MHz
(signal field) or 657~MHz (Ratos field) was measured
using a spectrum analyzer in the zero span mode, with a temporal
resolution of 200~ns.

We performed a set of preliminary measurements to verify the functionality of our setup. First, with all three input fields continuously on, we observed generation of a field on the Ratos transition  due to four-wave mixing. Second, we checked that the signal field experienced EIT when only the pump was present and observed slowdown of the 400-ns signal pulse. Third, we performed a ``classic'' storage of light experiment \cite{lightstorage} by turning the pump field off after the signal pulse entered the cell, and turning it back on after a few hundred $\mu$s. Fourth, we stored the signal pulse and then recovered it on the Ratos transition using the retrieve field akin to Zibrov and co-workers \cite{Zibrov:02}. Finally, by reducing the delay $\Delta t$ between the turn-off of the pump and the turn-on of the retrieve field to small negative values, we observed RATOS [Fig.~2].

Fig.~\ref{fig:pulses} shows the observed waveforms with the values of $\Delta \tau$ varied between -??? and ??? $\mu$s. In contrast to optical storage ($\Delta t>0$), the position of the retrieved pulse with negative values of the delay does not depend on $\Delta t$, which shows the adiabatic character of RATOS.

In classical four-wave mixing, the electric field of the created mode is
proportional to the electric field of the three mixing fields: $\langle\hat a_2\rangle \propto \Omega_1\Omega_2\langle\hat
    a_1\rangle.$ In contrast, the adiabatic transfer process studied here should not depend on the exact parameters of the control fields, as long as the switching takes place while the signal pulse is inside the cell. We verified the adiabatic nature of our experiment by varying the intensity of the retrieve field and monitoring the shape of the output Ratos pulse.

The result of this measurement is shown in Fig.~\ref{fig:powerdependence}. As expected, the time integrated intensity of the Ratos pulse shows only a weak dependence of the created field mode for sufficiently high retrieve field intensities. With the retrieve field below 2 mW, the EIT effect was not sufficiently pronounced so the Ratos pulse experienced partial absorption. The residual
dependence for higher powers is due to a Gaussian geometric
profile of the control fields, so the EIT was poor in the wings of the
beams.

The Ratos pulse \emph{shape}, on the other hand, depends strongly on the control field parameters. As evidenced by Fig.~\ref{fig:powerdependence}, the peak intensity and the temporal width of the Ratos pulse are, respectively, proportional and inversely proportional to the power of the retrieve laser $\Omega_2$. This is because the group velocity of the signal pulse in an EIT medium is proportional to the width of the EIT resonance \cite{fleischhauer:633}. The latter, in a Doppler-broadened, weakly decohering
medium is essentially proportional to the control field intensity \cite{Figueroa:06}.

\begin{figure}
\centerline{\includegraphics[width=0.9\columnwidth]{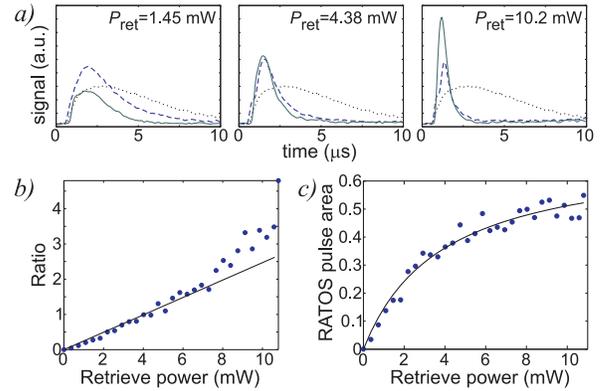}}
  \caption{\small Beam splitting via RATOS. a) Example waveforms for different retrieve pulse powers $P_{\rm ret}$. The Ratos field ($\hat a_2$) is shown with a solid line, the transmitted signal ($\hat a_1$) with a dashed line. The dotted line displays the transmitted, slowed down signal pulse in the absence of the retrieve field (regular EIT).
(b) Energy ratio of the Ratos pulse and the transmitted signal pulse, as a function of the retrieve field power. (c) Energy of the restored Ratos pulse normalized to the energy of the
  slowed down pulse; the solid line is a theoretical fit.}\label{splitpulse}
\end{figure}

In order to realize an optically controlled beam splitter, the pump field
$\Omega_1$ was kept on continuously, while $\Omega_2$ was turned
on when the signal pulse had fully entered the Rubidium cell. In
this case the quantum state of the signal mode $\hat a_1$ was
transferred into a superposition $\hat b$ of the modes $\hat a_i$,
given by Eq.~(\ref{superposition}). The power of the pump field was kept constant
at 4~mW, and the signal fields $\hat a_1$ and $\hat
a_2$ were measured for different powers of the retrieve field.
The results of this experiment are shown in Fig.~\ref{splitpulse}.
The three waveforms of the output signal and Ratos fields at different retrieve field powers [Fig.~\ref{splitpulse}(a)] illustrate the dynamics of multimode dark-state polaritons in the EIT medium. Even though the pump field remains the same in all three plots, the group velocity of the
signal pulse (which couples to the pump field through an excited level) increases with the intensity of the retrieve field. This happens because in the presence of the retrieve field, the signal is no longer an independent EIT mode, but a part of a multimode dark-state polariton whose group velocity is proportional to the weighted quadratic mean of all the control Rabi frequencies \cite{appel:06}.

Fig.~\ref{splitpulse}(b) displays the energy ratio of the Ratos and output signal fields as a function of the retrieve intensity. The observed proportional dependence is explained by Eqs.~(\ref{superposition}) and (\ref{splittingratio}): the fraction of a particular signal field in a multimode polariton is proportional to the Rabi frequency of the associated control field. The observed deviation from the linear fit is due to a systematic error in evaluating the output signal energy: at high retrieve powers, the signal is small so the relative error increases.


If the energy of the
retrieved pulse 
is plotted against the power of the retrieve field, a
dependence given by $ P_{\rm Ratos} = P_{\text{signal}}P_{\rm pump}/(c P_{\rm pump} +P_{\rm ret})$ is expected, where $c$ is a constant depending on the oscillator strengths of
the transitions used and the beam radii.
In  Fig.~\ref{splitpulse}(c), this function
is fitted to the data, with an overall coefficient added to account for the efficiency of the process. The fit yields a transfer efficiency of $a$=70\% with respect to the slowed down pulse.


In summary, we have demonstrated the possibility for adiabatic frequency conversion and routing of
optical information carried by light between two signal modes in a multi-$\Lambda$ EIT configuration.  
We foresee that RATOS will be useful in a variety of quantum commumnication and quantum engineering applications. Of special interest is the extension of RATOS to solid state systems \cite{turukhin:02}, where the level
structure allows access to nearly arbitrary frequencies for the
created Ratos pulse.

The measurements have been performed using classical light pulses,
thus the efficient transfer of the quantum state is yet to be
demonstrated. To this end we have set up a narrowband parametric quantum light
source\cite{Dallas},
which can be used to verify conservation of a quantum state
during the transfer.


This work was supported by NSERC, CIAR, AIF, CFI and Quantum\emph{Works}. We appreciate helpful discussions with K.-P. Marzlin.

\end{document}